\RequirePackage{fix-cm}
\RequirePackage{amsmath}
\documentclass[twocolumn,epjc3]{svjour3}  

\smartqed  % flush right qed marks, e.g. at end of proof
\RequirePackage{graphicx}
\RequirePackage{mathptmx}      % use Times fonts if available on your TeX 
\usepackage{color}
\usepackage{mathrsfs}
\usepackage{amssymb, bm}
\usepackage{url}

\newcommand{\be}{\begin{equation}}
\newcommand{\ee}{\end{equation}}

\begin{document}

\title{Non-geodesic timelike observers and the ultralocal limit
  %\thanksref{t1}
}
%\subtitle{}


\author{Reza Saadati\thanksref{e1,addr1}
	\and
Luca Valsan\thanksref{e2,addr2}
	\and
Valerio Faraoni \thanksref{e3,addr1} 
}

\thankstext{e1}{e-mail: rsaadati@ubishops.ca}
\thankstext{e2}{e-mail: luca.valsan@mcgill.ca}
\thankstext{e3}{e-mail: vfaraoni@ubishops.ca}


\institute{Department of Physics \& Astronomy, Bishop's University, 
2600 College Street, Sherbrooke, Qu\'ebec, Canada J1M~1Z7 \label{addr1}
\and
Department of Physics, McGill University, 
3600 rue University, 
Montr\'eal, Quebec
Canada H3A 2T8  \label{addr2}
}

%\institute{Department of Physics \& Astronomy, Bishop's University, 
%2600 College Street, Sherbrooke, Qu\'ebec, Canada J1M~1Z7 \label{addr3}
%}

\date{Received: date / Accepted: date}

\maketitle

\begin{abstract}

The ultralocal limit along timelike geodesics, in which any geometry 
reduces to Bianchi~I, does not extend to non-geodesic 
timelike observers. Exceptions are discussed, including particles with 
variable mass, test particles in Einstein frame scalar-tensor gravity, 
and self-interacting dark matter.

\end{abstract}

\section{Introduction}
\label{sec:1}
\setcounter{equation}{0}

Ultralocal limits in general relativity (GR) have been the subject of 
renewed interest in recent years. In an ultralocal limit, one restricts 
oneself to looking at the 3-space in an infinitesimally small neighborhood 
of a null or timelike geodesic. In physical terms, in an ultralocal limit 
in which the light cones close and collapse on the observer's wordline, 
nothing propagates in its 3-space. This situation is often pictured as 
saying that in this limit the speed of light $c$ tends to zero. This is 
the Carrollian limit of GR 
\cite{LevyLeblond65,Bacry:1968zf,Dautcourt:1997hb}, the opposite situation 
to Newtonian physics in which the light cones instead open up and flatten 
out on the 3-space of the observer, while $c$ tends to infinity. The 
ultra-relativistic Carroll limit, originally studied as a formal limit of 
the Poincar\'e algebra, is associated with quantum effects in strong 
gravity regimes and the AdS/CFT correspondence (e.g., 
\cite{Bergshoeff:2017btm,Duval:2014uoa,Hansen}). While, in reality, $c$ is 
a constant of nature with a definite value, considering these limits makes 
sense physically.

Almost half a century ago, Penrose demonstrated that along null 
geodesics every spacetime 
metric reduces to a plane wave in the ultralocal limit \cite{Penrose}. 
This work was related to the importance of $pp$-waves \cite{Blau} 
and can be interpreted physically by saying that, for a freely falling 
observer reaching asymptotically the speed of light, any gravitational 
field would look that of an exact plane gravitational wave passing by 
at light speed. An important feature of this result is its universality.

Another known application of the ultralocal limit is the study 
of the approach to spacelike singularities: Belinski, Khalatnikov and 
Lifshitz \cite{Belinsky:1970ew,Belinski:2017fas} showed that in the 
ultra-local limit, the vacuum geometry near a spacelike singularity is 
approximated by a Kasner one with line element
\be
ds^2 =- dt^2  
+ \left( a_1 t^{p_1} \right)^2 dx^2  
+ \left( a_2 t^{p_2} \right)^2 dy^2  
+ \left( a_3 t^{p_3} \right)^2 dz^2  
\ee
with $ p_1 + p_2 + p_3=1 $, $ p_1^2 + p_2^2 + p_3^2 =1$ and constant 
$a_i$ ~($i=1,2,3$). This is a 
special case of a Bianchi~I geometry, with the exponents $p_i$ depending 
on the spatial coordinates, but this dependence vanishes in the ultralocal 
limit.

More recently, Cropp and Visser \cite{Cropp:2010yj,Cropp:2011er} worked 
out a similar limit for geodesic {\em timelike} observers, which is a more 
physical situation. They demonstrated that freely falling observers see 
{\em any} gravitational field as a Bianchi~I model in the ultralocal limit 
\cite{Cropp:2010yj,Cropp:2011er}. Here the Cropp-Visser ultralocal limit 
is revisited and it is shown that timelike {\em non-geodesic} observers do 
not enjoy such a universal ultralocal limit, but that there are certain 
physically relevant exceptions to this rule.

In general, the Cropp-Visser ultralocal limit fails for non-geodesic 
timelike curves because the very first step, the introduction of Gaussian 
normal (or ``synchronous'') coordinates, is not possible along 
non-geodesic curves.  However, there are very special and restricted 
situations in which an accelerated particle is subject to a four-force but 
the discussion of \cite{Cropp:2010yj,Cropp:2011er} still applies in the 
ultralocal limit and the geometry looks again like that of a Bianchi~I 
universe to this particle. This situation occurs when the four-force 
acting on the observer is tangential to its trajectory (i.e., parallel or 
anti-parallel to its four-tangent $u^\mu$), in which case synchronous 
coordinates can still be introduced and the derivation of the ultralocal 
limit in \cite{Cropp:2010yj,Cropp:2011er} proceeds as for geodesic 
timelike curves.

In the following sections we recall the standard derivation of the 
ultralocal limit of \cite{Cropp:2010yj,Cropp:2011er} (Sec.~\ref{sec:2}),  
showing 
explicitly why the 
procedure fails for non-geodesic curves (Sec.~\ref{sec:3}); we then 
extend the validity of the Cropp-Visser limit  
to non-geodesic timelike curves with four-force parallel to the curve 
(Sec.~\ref{sec:4}). Section~\ref{sec:5} muses about generic 
accelerated observers, while Sec.~\ref{sec:6} summarizes the results.

We follow the notation of Ref.~\cite{Wald:1984rg}. Units are used in which 
the speed of light $c$ and Newton's constant $G$ are unity. Greek 
(spacetime) indices run from 0 to 3, while (purely spatial) Latin indices 
run from 1 to 3, $g_{\mu\nu}$ denotes the spacetime metric and 
$\nabla_{\alpha} $ the corresponding covariant derivative.

\section{The ultralocal limit for timelike geodesics}
\label{sec:2}

We report briefly the Cropp-Visser derivation \cite{Cropp:2010yj}.  One 
begins by assuming an observer 
freely 
falling along a timelike 
geodesic $\gamma$ of the spacetime metric $g_{\mu\nu}( x^{\alpha})$. The 
first step consists of adopting synchronous (or Gaussian normal) 
coordinates,  which are always defined locally in the neighborhood of a 
timelike geodesic curve. Locate the origin of the spatial 
coordinates $x^i$ at the curve  $\gamma$. In these coordinates, the line 
element assumes the form
\be
ds^2=-c^2dt^2 + g_{ij}\left( t, x^k \right) dx^i dx^j \,,
\ee
i.e., $g_{00}=-1$ and $g_{0i}=0$ in these coordinates.  Now consider the 
coordinate transformation  
\be
x^i\to x^{i'}( x^j) = \epsilon \, x^i \,, 
\ee
where $\epsilon $ is a positive constant, which involves only a 
rescaling of the spatial coordinates. 
We will be interested in the limit $ \epsilon \to 0$, which clearly 
corresponds from the outset  to shrinking the 3-space around the 
worldline $\gamma$ onto the worldline itself. The line element has 
taken the form 
\be
ds^2= -c^2dt^2 + \epsilon^2 g_{ij}\left( t, \epsilon \, x^k \right) 
dx^i dx^j \,;
\ee
now rescale the speed of light as $c\to \epsilon \, c$, obtaining  
\be
ds^2=\epsilon^2 \left[ -c^2dt^2 + g_{ij}\left( t, \epsilon \, x^k \right) 
dx^i dx^j \right] \,,
\ee
and perform the conformal transformation $ ds^2 \to d\tilde{s}^2 = 
\epsilon^{-2} ds^2$ (as is done also for the Penrose limit along null 
geodesics \cite{Penrose}), which gives
\be
ds^2=  -c^2dt^2 + g_{ij}\left( t, \epsilon \, x^k \right) 
dx^i dx^j .
\ee
Finally, we take the limit $\epsilon \to 0$, obtaining the line element 
\be
ds^2=  -c^2dt^2 + g_{ij}\left( t, \vec{0} \right) dx^i dx^j 
\,,\label{BianchiI}
\ee
which has the form describing a Bianchi~I universe 
\cite{EMMacC,ExactSolutions}.

\section{The failure of synchronous coordinates and of the universal  
ultralocal limit}
\label{sec:3}

Let us ask now the question of how the spacetime geometry will appear to 
an {\em accelerated} timelike observer. Unfortunately, the derivation of 
the result that it appears as a Bianchi~I universe fails at its very first 
step: synchronous coordinates cannot be introduced along non-geodesic 
timelike curves. To see why, let us examine how the standard derivation of 
Gaussian normal coordinates (e.g., \cite{Wald:1984rg}) is modified by the 
fact that these curves deviate from geodesics, and what the obstruction is 
precisely.

Assume that a massive particle is subject to a four-force per unit mass  
$f^{\mu}$ and follows the spacetime trajectory $\gamma'$ with timelike 
tangent $u^{\mu}$ described by the equation
\be
u^{\nu} \nabla_{\nu} u^{\mu}= \frac{du^{\mu}}{d\tau} + 
\Gamma^{\mu}_{\alpha\beta} \, u^{\alpha} u^{\beta} =f^{\mu} \,,
\label{non-geodesic}
\ee
where $\tau$ is the proper time along the trajectory and 
$\Gamma^{\mu}_{\alpha\beta}$  
are the Christoffel symbols of the connection. To introduce synchronous  
coordinates $\left( \tau, x^i \right)$, the wordline $\gamma'$ of the 
particle must be normal 
to every  hypersurface of 3-space $\Sigma_{\tau}$ of constant time 
$\tau$, in which there are spatial coordinates $x^i$ with three purely 
spatial coordinate vectors $X^{\nu}$ according to the observer $u^\mu$.  
That is, for all  $\Sigma_{\tau}$, we must have $ u^{\mu}  X_{\mu} =0$ 
for each of the three vectors $X^{\alpha}$. 
This means that, defining the spatial coordinates so that $X^{\mu} 
u_{\mu}=0$ 
initially, this condition is preserved along $\gamma'$, i.e., 
\be
\frac{D}{D\tau} \left( X^{\mu} u_{\mu} \right) = u^{\nu} \nabla_{\nu} 
\left( X_{\mu} u^{\mu} \right)=0 \,,
\ee
but this property fails to hold because it is instead
\be
u^{\nu} \nabla_{\nu} \left( X_{\mu} u^{\mu} \right)= X_{\alpha} f^{\alpha} 
\,.
\ee
To wit, 
\begin{eqnarray}
u^{\nu} \nabla_{\nu} \left( X_{\alpha} u^{\alpha} \right) &=& u_{\alpha} 
u^{\beta} \nabla_{\beta} X^{\alpha} +X^{\alpha} u^{\beta} 
\nabla_{\beta} u^{\alpha} \nonumber\\
&&\nonumber\\
&=& u_{\alpha} u^{\beta} \nabla_{\beta} X^{\alpha} +X_{\alpha} f^{\alpha} 
\,.
\end{eqnarray}
Since $u^{\mu}$ and $X^{\mu}$are the vectors of a coordinate basis on the 
spacetime manifold their commutator vanishes, yielding 
\be
u^{\beta} \nabla_{\beta} \left( X_{\alpha} u^{\alpha} \right) =  
u_{\alpha} X^{\beta} \nabla_{\beta} u^{\alpha} +X_{\alpha} f^{\alpha} = 
X_{\alpha} f^{\alpha} \label{waldeqn}
\ee
because the normalization $u^{\mu} u_{\mu}=-1$ implies that $ u^{\alpha} 
\nabla_{\beta} u_{\alpha}=0$.

The right-hand side of Eq.~(\ref{waldeqn}) vanishes if and only if the 
particle is free or is subject to a four-force perpendicular to $X^{\mu}$. 
A clear obstruction is thus identified in constructing synchronous 
coordinates along non-geodesic timelike curves.

\section{Exceptions}
\label{sec:4}

 If the four-force $f^{\mu}$ is parallel or antiparallel to the 
four-velocity $u^{\alpha}$, the product $X_{\alpha} f^{\alpha}$ vanishes, 
Eq.~(\ref{waldeqn}) reduces to the affinely parametrized geodesic 
equation, and the obstruction to constructing synchronous coordinates is 
removed.

The condition that the four-force acting on a massive particle be 
(anti-)parallel to the particle trajectory's tangent seems rather 
exceptional, to the point that is not contemplated in most textbooks. 
However, this circumstance is far from unphysical, including situations in 
which the particle mass changes along the trajectory, the ultralocal limit 
of particles subject only to gravity in the Einstein frame description of 
scalar-tensor gravity \cite{Faraoni:2020ejh}, fluid elements of perfect or 
imperfect fluids in Friedmann-Lema\^itre-Robertson-Walker (FLRW) or 
Bianchi universes, mass-changing particles in cosmology and in 
scalar-tensor gravity 
\cite{Mbelek:1998vu,Mbelek:2004ff,Damour:1990tw,Casas:1991ky,Garcia-Bellido:1992xlz,AndersonCarroll97}, 
and certain scenarios in which self-interacting dark matter is effectively 
subject to a sort of anti-friction \cite{Zimdahl:2000zm}. In the standard 
textbook treatment, the mass $m$ of a particle following a timelike 
worldline $\gamma'$ with four-tangent $u^\mu$ is constant and the 
four-force acting on it is simply $f^{\alpha}=m a^{\alpha}$, where 
$a^{\beta} \equiv \dot{u}^{\beta} \equiv u^{\mu} \nabla_{\mu} u^{\beta} $ 
is the particle's four-acceleration. Rockets are prototypical systems with 
variable mass and exact solutions describing rockets 
\cite{Kinnersley:1969zz,Kinnersley:1970zw,Bonnor:1994ir,Damour:1994dy,Dain:1996eq,Bonnor:1997jc,Podolsky:2008wn,Podolsky:2010ck,Ge:2011ey,Hogan:2020duo,ExactSolutions} 
and solar sails  \cite{Forward84,Fuzfa:2019djg,Fuzfa:2020dgw} abound in 
GR.

Another instance of four-force parallel to a particle trajectory arises 
in cosmology. Quantum processes in the early universe lead to 
particle production and 
cause a  negative bulk pressure  
\cite{Zeldovich:1970si,Hu}. This mechanism could potentially  drive 
inflation, as suggested in 
\cite{Zimdahl:2000zm,Schwarz:2001cf,Zimdahl:1996cg,Zimdahl:1997qe,ZimdahlBalakin98a,Zimdahl:1998zq}.  
Likewise, the self- interaction of dark matter can cause negative bulk 
stresses, a mechanism currently under investigation as a possible cause  
of the present acceleration of the universe \cite{Zimdahl:2000zm}. 
Indeed, this self-interaction would be responsible for a cosmic 
``antifriction'' on the dark matter fluid, i.e., for a force antiparallel 
to the timelike four-trajectories of dark matter particles 
\cite{Zimdahl:2000zm}.

Yet again, in (Jordan frame) scalar-tensor gravity 
\cite{Jordan38,Jordan:1959eg,Brans:1961sx,Bergmann:1968ve,Nordtvedt:1968qs,Wagoner:1970vr,Nordtvedt:1970uv}, a scalar 
field degree of freedom $\phi$ (the Brans-Dicke-like scalar) appears 
together with the two 
massless spin two modes contained in the metric tensor in GR. The action 
is
\begin{eqnarray}
S_\mathrm{ST} & = & \frac{1}{16\pi} \int d^4 x \, \sqrt{-g} \, \left[ \phi R 
-\frac{\omega(\phi)}{\phi} \, \nabla^{\mu} \phi \nabla_{\mu} \phi 
-V(\phi) \right. \nonumber\\
&&\nonumber\\
&\, & \left. +{\cal L}^\mathrm{(m)} \right] \,, \label{STaction}
\end{eqnarray}
where $g$ is the metric determinant, $R$ is the Ricci scalar, $\omega 
(\phi)$ is the Brans-Dicke coupling, $V(\phi)$ is a 
potential for the scalar field, and ${\cal L}^\mathrm{(m)}$ is the matter 
Lagrangian density.  
The Jordan frame variables are $\left( 
g_{\mu\nu}, \phi \right)$, Newton's constant $G$ is replaced 
by the effective gravitational coupling strength 
$G_\mathrm{eff}\simeq 1/\phi$, and $\phi$ couples directly to 
the Ricci scalar $R$. In the Einstein frame representation of 
these theories, the variables $\left( \tilde{g}_{\mu\nu}, \tilde{\phi} 
\right)$ are defined by
\be
\tilde{g}_{\mu\nu} = \phi \, g_{\mu\nu} \,,
\ee
\be
d\tilde{\phi} =\sqrt{ \frac{2\omega+3}{16\pi }} \, 
\frac{d\phi}{\phi} \,.
\ee
In this frame the scalar $\tilde{\phi}$ does not couple 
explicitly to the spacetime curvature, but couples directly to 
matter and the action~(\ref{STaction}) is rewritten as 
\begin{eqnarray}
S_\mathrm{ST}&=&\int d^4 x \, \sqrt{-\tilde{g}} \, \left[ \frac{ 
\tilde{R}}{16\pi}   
-\frac{1}{2} \, \tilde{g}^{\mu\nu} \tilde{\nabla}_{\mu} \tilde{\phi} 
\tilde{\nabla}_{\nu} \tilde{\phi}  
-U( \tilde{\phi}) \right.\nonumber\\
&&\nonumber\\
&\, & \left. + \frac{{\cal L}^\mathrm{(m)}}{ 
\phi^2(\tilde{\phi})}  \right]  
\end{eqnarray}
in terms of the Einstein frame variables $\left( \tilde{g}_{\mu\nu} 
\tilde{\phi} \right)$, where 
\be
U \left( \tilde{\phi}\right) = \frac{ V(\phi)}{ \phi^2} \Big|_{\phi=\phi( 
\tilde{\phi}) }.
\ee 
As a consequence particles subject only to gravity, 
which follow geodesics in the Jordan frame, deviate from 
geodesics in the Einstein frame, according to 
\cite{Dicke:1961gz,Faraoni:2004pi} 
\be
\frac{d^2 x^{\mu}}{d\tau^2} + \tilde{\Gamma}^{\mu}_{\alpha\beta} \, 
\frac{dx^{\alpha}}{d\tau} \,  
\frac{dx^{\beta}}{d\tau}   
= \sqrt{\frac{4\pi}{2\omega+3}} \, \tilde{\nabla}^{\mu} \tilde{\phi}  
\,.
\ee 
The common interpretation of this equation is that the mass 
of a test particle (which was constant in the Jordan frame) 
instead depends on $\tilde{\phi}$ in the Einstein frame and 
is no longer geodesic.\footnote{However, since the scalar 
field $\phi$ has gravitational nature, this particle is still 
subject only to gravity and to no other forces, but now 
gravity is described by {\em both} $g_{\mu\nu}$ and $\phi$.} The 
gradient of $\phi$ across spacetime translates in the 
dependence of the particle mass $m$ on the spacetime position 
and in a fifth force proportional to $ \tilde{\nabla}^{\mu} \tilde{\phi}$  
\cite{Faraoni:2004pi,Faraoni:2020ejh}.\footnote{In dilaton 
gravity and in the low-energy limit of string theories, a 
similar equation appears but  the coupling of the dilaton 
field to Standard Model particles is not universal, usually 
depending on their quark content 
\cite{Taylor:1988nw,Damour:1994zq,Gasperini:1999ne}.} 

Once the particle trajectory $\gamma'$ is fixed by initial conditions, 
the particle mass depends only on the proper time $\tau$ 
along this trajectory. In the ultralocal limit, the spatial 
dependence of $\phi$ is killed anyway, leaving $\phi \left(t, 
\vec{0} \right)$ instead of $\phi \left(t, \vec{x} \right)$, 
and $m=m(\tau)$. In these conditions, one effectively has a 
time-dependent mass along the trajectory and a four-force 
$f^\mu$ parallel to its four-tangent $u^\mu$. The conclusion that 
test particles subject only to gravity ``see'' ultra-locally 
any spacetime geometry as a Bianchi~I geometry, therefore, 
holds for Einstein frame scalar-tensor gravity.

Another GR situation in which a four-force parallel to 
massive particle worldlines  occurs in cosmology. Consider a FLRW 
universe sourced by a fluid, with line element
\be
ds^2 =-dt^2 +a^2(t) \left[ \frac{dr^2}{1-kr^2} +r^2 \left( d\vartheta^2 
+\sin^2\vartheta \, d\varphi^2 \right)\right]
\ee
in comoving coordinates $\left( t, r, \vartheta, \varphi \right)$. Except 
for the case in which the matter fluid is dust or a cosmological constant 
$\Lambda$ (corresponding to stress-energy tensor 
$T_{\mu\nu}^{(\Lambda)}=-\frac{\Lambda}{8\pi} \, g_{\mu\nu}$), this fluid 
has pressure $P(t)$ and pressure gradient $\nabla_{\mu} P \neq 0$, which 
generates a four-force pointing in the (comoving) time direction $u^\mu$. 
In fact, the four-force and four-acceleration must have vanishing spatial 
components to respect spatial isotropy. As a consequence, fluid particles 
are accelerated and deviate from geodesics. The fluid particles obey the 
equation (e.g., \cite{Faraoni:2020ejh})
\be
\frac{d^2 x^{\mu}}{dt^2} + \Gamma^{\mu}_{\alpha\beta} \, 
\frac{dx^{\alpha}}{dt} \, \frac{ dx^{\beta}}{dt}  
= B \, \frac{dx^{\mu}}{dt} \,, \label{non-affinegeodesic}
\ee
where $B$ depends on the position along the particle worldline. 
This equation is just  the non-affinely parametrized geodesic 
equation, where the proper time $t$ of comoving observers is not an 
affine parameter. It is always 
possible to 
switch from $t$  to an affine parameter: then the 
right-hand side of the geodesic equation~(\ref{non-affinegeodesic})  
vanishes. We conclude that this 
four-acceleration is somehow trivial, but the reparametrization is 
not.  If $s$ is an affine 
parameter, the function $B$ in Eq.~(\ref{non-affinegeodesic}) 
is $B(t)= \frac{dt}{ds} \, \frac{d^2s}{dt^2} $ 
(\cite{EMMacC,Faraoni:2020ejh}, see Appendix~A of 
\cite{Faraoni:2023hwu}).  
The equation describing the worldlines  
of the fluid elements cannot be affinely parameterized by 
the 
fluid's proper time, which causes a four-force 
parallel to the four-velocity $u^\mu$  
\cite{Faraoni:2020ejh,Faraoni:2023hwu}.  From the purely mathematical 
point of view, this fact is immaterial but the difference 
between the proper time of the comoving observers  and an affine parameter  
matters from the physical point of view. FLRW cosmology is always 
described using the frame of 
comoving observers. 

This discussion extends to Bianchi cosmologies, reaching the conclusion 
that pressure gradients and anisotropic stresses generate four-forces 
parallel to the fluid worldlines, therefore the ultralocal limit applies 
\cite{Faraoni:2023hwu}. This fact can perhaps be taken as a consistency 
check of the ultralocal limit for non-geodesic fluids, but is almost  
trivial since it states that a FLRW or Bianchi universe (possibly with 
spatial curvature)  
 looks locally like a Bianchi~I spacetime! FLRW is a special case of 
Bianchi~I with vanishing spatial anisotropy and any spatially curved 
Bianchi model reduces to Bianchi~I when the spatial curvature is 
negligible in the local limit.

\section{General accelerated observer}
\label{sec:5}

Since, in the general case, one cannot introduce synchronous coordinates 
along the wordline $\gamma'$ of an accelerated observer, the best on can 
do to perform an ultralocal limit is the following. Given the 
force $f^\mu$ acting on an accelerated timelike observer and it worldline 
$\gamma'$ satisfying Eq.~(\ref{non-geodesic}), one can perform a  
rescaling 
of the spatial  coordinates in the 3-space with Riemannian metric 
$h_{\mu\nu}= 
g_{\mu\nu} + u_{\mu} u_{\nu} $ and then take the limit $\epsilon\to0$ as 
described in 
Sec.~\ref{sec:2}, obtaining
\begin{eqnarray}
ds^2 &=& g_{00} \left( t(\tau) , \vec{0} \right) dt^2 
+ 2 g_{0i} \left( t(\tau) , \vec{0} \right) dt dx^i \nonumber\\
&&\nonumber\\
&\, & + g_{ij} \left( t(\tau) , \vec{0} \right) dx^i dx^j \,,
\end{eqnarray}
where $t(\tau)=x^0(\tau)$ is the time component of the solution 
$x^{\mu}(\tau)$ of Eq.~(\ref{non-geodesic}). One can then redefine the 
time 
coordinate $t \to \bar{t} $ according to  $
-g_{00} \left( t(\tau) , \vec{0} \right) dt^2 =d\bar{t}^2 $, 
or
\be
\bar{t}= \int d\tau \, \frac{dt}{d\tau}\,  \sqrt{-g_{00}\left( t(\tau) , 
\vec{0} \right)} \,.\label{accipicchia}
\ee
$d\bar{t}$ is an exact differential because the integrand in 
Eq.~(\ref{accipicchia}) depends only on $\tau$. We are left with the 
time-dependent geometry 
\be
ds^2 = - d\bar{t}^2 
+ 2 \bar{g}_{0i} \left( \bar{t} \right) d\bar{t} dx^i 
+ g_{ij} \left( \bar{t} \right) dx^i dx^j \,, \label{questa*}
\ee
where
$ \bar{g}_{0i} \left( \bar{t} \right) = g_{0i} \left( t(\tau(\bar{t}) ) 
, \vec{0}  \right)$, $ 
\bar{g}_{ii} \left( \bar{t} \right) = g_{ij} \left( t(\tau(\bar{t}) ) 
, \vec{0}  \right)$ are obtained by inverting the relation 
$\bar{t}(\tau)$. This geometry is still very general and the  
universality of the ultralocal limit for geodesic observers (in which {\em 
any} geometry looks like Bianchi I) is completely lost.

One could pose the problem of whether a prescribed 
geometry~(\ref{questa*})  can be obtained by tailoring the four-force 
$f^\mu \left( x^{\alpha}, \dot{x}^{\alpha} \right)$ acting on an 
accelerated observer. However, this problem is not amenable to 
mathematical treatment. Even if Eq.~(\ref{non-geodesic}) could be solved 
analytically and explicitly for all forces $f^\mu$ deemed necessary to 
achive a required geometry $\bar{g}_{\mu\nu}(\bar{t})$ (which is in 
practice impossible), the solution would comprise only four functions 
$x^{\mu} (\tau)$ while specifying completely the metric components 
$\bar{g}_{0i}, \bar{g}_{ij} $ requires one to impose nine conditions. 
There 
is no precise mathematical way to relate these nine functional relations 
(or a subset of four of them) to the force $f^\mu$, and no way to pose a 
mathematically well-defined problem. It may well be that in some special 
cases a solution to this problem exists: for example, thanks to 
Ref.~\cite{Cropp:2010yj} we know that the problem of imposing 
$\bar{g}_{\mu\nu}$ to be of the Bianchi~I form has a solution for 
$f^{\mu}=0$ (we do not know whether this is the unique solution). However, 
in general, no solution to this problem posed vaguely in mathematical 
terms can be expected.

\section{Conclusions}
\label{sec:6}

The ultralocal limit along timelike geodesics, in which all geometries 
look Bianchi~I to freely falling observers, does not extend to the 
non-geodesic timelike curves followed by accelerated particles, with the 
exception of particles subject to a four-force parallel (or antiparallel) 
to the four-tangent to the trajectory. Although quite special, this 
situation should not be dismissed {\em a priori} as unphysical or 
irrelevant.  As already noted, the case of particles subject to pressure 
gradients in cosmology is trivial because one starts with cosmology and 
recovers a cosmology in the ultralocal limit, but other situations are by 
all means not trivial. They include particles with variable mass (e.g., 
rockets and solar sails, which are the subject of a non-negligible 
literature in GR 
\cite{Kinnersley:1969zz,Kinnersley:1970zw,Bonnor:1994ir,Damour:1994dy,Dain:1996eq,Bonnor:1997jc,Podolsky:2008wn,Podolsky:2010ck,Ge:2011ey,Hogan:2020duo,ExactSolutions,Forward84,Fuzfa:2019djg,Fuzfa:2020dgw}), 
test particles in Einstein frame scalar-tensor gravity, and 
self-interacting dark matter particles in certain scenarios 
\cite{Zimdahl:2000zm}. The fact that the ultralocal limit extends to 
scalar-tensor gravity comes to no surprise since the field equations are 
not used in the derivation of the ultralocal limits 
\cite{Penrose,Cropp:2010yj,Cropp:2011er}.

We have in mind an application of the ultralocal limit to a new 
thermal view of scalar-tensor (including ``viable'' Horndeski) gravity 
\cite{Faraoni:2018qdr,Faraoni:2021lfc,Faraoni:2021jri,Giusti:2022tgq,Faraoni:2023hwu,Quiros:2019gai,Giusti:2021sku,Miranda:2022wkz,Giardino:2022sdv,Faraoni:2022gry,Miranda:2022uyk,Faraoni:2022doe,Faraoni:2022jyd,Giardino:2023qlu,Faraoni:2022fxo,Giardino:2023sw,Miranda:2024dhw,Giardino:2023ygc,Karolinski:2024nwp,Houle:2024sxs,Giardino:2024pla} 
in the case in which the gradient $\nabla^{\mu} \phi$ of the gravitational 
scalar field is timelike and future-oriented. In this formalism, the 
effective stress-energy tensor of $\phi$ has the form of a dissipative 
fluid stress-energy tensor and satisfies Eckart's \cite{Eckart40} 
generalization of the Fourier law of heat conduction, which leads one to 
defining a ``temperature of gravity''. Then, the approach of scalar-tensor 
gravity to general relativity is analogous to the dissipation of heat in a 
medium; the relaxation of gravity to general relativity is described as 
the approach of a heated medium to its zero-temperature equilibrium state. 
This new formalism is still under development, but Bianchi~I models have 
been discussed in this context \cite{Houle:2024sxs}. The fact that in the 
ultralocal limit all scalar-tensor spacetimes are seen as Bianchi~I, and 
the reduction of Einstein frame non-geodesic observers to geodesic ones 
discussed in this work, has several potential implications for the new 
thermal view of scalar-tensor gravity.  Among them, we mention the 
following: it is possible that non-viable Horndeski gravity (i.e., the 
class of theories in which the speed of gravitational waves is not equal 
to the speed of light), which could not be included in the thermal view, 
becomes tractable in the ultralocal limit because of the simplifications 
that this limit brings to the effective fluid quantities. This application 
will be discussed in future work.

\begin{acknowledgements}

We thank a referee for helpful comments leading to improvements in 
the original manuscript. V.~F. is supported by the Natural Sciences \& 
Engineering Research Council of Canada (grant 2023-03234).

\end{acknowledgements}

\bigskip
%\appendix
%\section{}
%\label{appendix:A}
%\renewcommand{\theequation}{A.\arabic{equation}}


\begin{thebibliography}{}

\bibitem{LevyLeblond65} J.-M. Levy-Leblond, ``Une nouvelle limite 
non-relativiste du groupe de Poincar\'e, Annales de l’institut Henri 
Poincar\'e (A) Physique Th\'eorique 
3 (1965) no. 1, 1–12. http://eudml.org/doc/75509.

\bibitem{Bacry:1968zf}
H.~Bacry and J.~Levy-Leblond,
``Possible kinematics,''
J. Math. Phys. \textbf{9}, 1605-1614 (1968)
doi:10.1063/1.1664490

\bibitem{Dautcourt:1997hb}
G.~Dautcourt,
``On the ultrarelativistic limit of general relativity,''
Acta Phys. Polon. B \textbf{29}, 1047-1055 (1998)
[arXiv:gr-qc/9801093 [gr-qc]].

\bibitem{Bergshoeff:2017btm}
E.~Bergshoeff, J.~Gomis, B.~Rollier, J.~Rosseel and T.~ter Veldhuis,
``Carroll versus Galilei Gravity,''
JHEP \textbf{03}, 165 (2017)
doi:10.1007/JHEP03(2017)165
[arXiv:1701.06156 [hep-th]].

\bibitem{Duval:2014uoa}
C.~Duval, G.~W.~Gibbons, P.~A.~Horvathy and P.~M.~Zhang,
``Carroll versus Newton and Galilei: two dual non-Einsteinian concepts of 
time,''
Class. Quant. Grav. \textbf{31}, 085016 (2014)
doi:10.1088/0264-9381/31/8/085016
[arXiv:1402.0657 [gr-qc]].

\bibitem{Hansen} D. Hansen, N. A. Obers, G. Oling and B. T. S\o gaard, 
"Carroll expansion of general relativity", SciPost Phys. {\bf 13}, 055 
(2022) doi:10.21468/SciPostPhys.13.3.055

\bibitem{Penrose} R. Penrose, ``Any space-time has a plane wave as a 
limit'', in {\em Differential Geometry and Relativity}, Mathematical 
Physics and Applied Mathematics, vol 3, edited by M. Cahen and M.Flato, M.
 (Springer, Dordrecht 1976). 
\url{https://doi.org/10.1007/978-94-010-1508-0_23}

\bibitem{Blau}
M. Blau, ``Plane Waves and Penrose Limits'', lectures notes of  
the 2004 Saalburg/Wolfersdorf Summer School, 
\url{http://www.blau.itp.unibe.ch/Lecturenotes.html} 
 
\bibitem{Belinsky:1970ew}
V.~A.~Belinsky, I.~M.~Khalatnikov and E.~M.~Lifshitz,
``Oscillatory approach to a singular point in the relativistic 
cosmology,''
Adv. Phys. \textbf{19}, 525-573 (1970)
doi:10.1080/00018737000101171

\bibitem{Belinski:2017fas}
V.~Belinski and M.~Henneaux,
{\it The Cosmological Singularity} 
(Cambridge University Press, Cambridge, 2017), 
doi:10.1017/9781107239333

\bibitem{Cropp:2010yj}
B.~Cropp and M.~Visser,
``Any spacetime has a Bianchi type I spacetime as a limit,''
Class. Quant. Grav. \textbf{28}, 055007 (2011)
doi:10.1088/0264-9381/28/5/055007
[arXiv:1008.4639 [gr-qc]].

\bibitem{Cropp:2011er} B.~Cropp, ``Applications of, and Extensions to, 
Selected Exact Solutions in General Relativity,'' [arXiv:1107.5618 
[gr-qc]].

\bibitem{Wald:1984rg} R.M.~Wald, {\em General Relativity} (Chicago 
University Press, Chicago, 1987), 
doi:10.7208/chicago/9780226870373.001.0001

\bibitem{EMMacC} G. F. R. Ellis, R. Maartens, and M. A. H. MacCallum, {\em 
Relativistic Cosmology} (Cambridge University Press, Cambridge, UK, 2012).

\bibitem{ExactSolutions} H. Stephani, D. Kramer, M. MacCallum, C. 
Hoenselaers, E. Herlt, {\em Exact Solutions of the Einstein Field 
Equations} (Cambridge University Press, Cambridge, UK, 2003).

\bibitem{Faraoni:2020ejh}
V.~Faraoni and G.~Vachon,
``Quasi-geodesics in relativistic gravity,''
Eur. Phys. J. C \textbf{81}, no.1, 22 (2021)
doi:10.1140/epjc/s10052-020-08808-9
[arXiv:2011.05891 [gr-qc]].

\bibitem{Mbelek:1998vu}
J.~P.~Mbelek,
``Motion of a test body in the presence of an external scalar field which 
respects the weak equivalence principle,''
Acta Cosmologica \textbf{24}, 127-148 (1998)
[arXiv:gr-qc/0402084 [gr-qc]].

\bibitem{Mbelek:2004ff}
J.~P.~Mbelek,
``Modelling the rotational curves of spiral galaxies with a scalar 
field,''
Astron. Astrophys. \textbf{424}, 761-764 (2004)
doi:10.1051/0004-6361:20040192
[arXiv:gr-qc/0411104 [gr-qc]].

\bibitem{Damour:1990tw}
T.~Damour, G.~W.~Gibbons and C.~Gundlach,
``Dark Matter, Time Varying $G$, and a Dilaton Field,''
Phys. Rev. Lett. \textbf{64}, 123-126 (1990)
doi:10.1103/PhysRevLett.64.123

\bibitem{Casas:1991ky}
J.~A.~Casas, J.~Garcia-Bellido and M.~Quiros,
``Scalar-tensor theories of gravity with phi dependent masses,''
Class. Quant. Grav. \textbf{9}, 1371-1384 (1992)
doi:10.1088/0264-9381/9/5/018
[arXiv:hep-ph/9204213 [hep-ph]].

\bibitem{Garcia-Bellido:1992xlz}
J.~Garcia-Bellido,
``Dark matter with variable masses,''
Int. J. Mod. Phys. D \textbf{2}, 85-95 (1993)
doi:10.1142/S0218271893000076
[arXiv:hep-ph/9205216 [hep-ph]].

\bibitem{AndersonCarroll97} G.~W. Anderson and S. Carroll, in {\em 
Proceedings COSMO-97}, 1st International Workshop on Particle Physics and 
the Early Universe, Ambleside, England, 1997, edited by L. Roszkowski 
(World Scientific, Singapore, 1997).

\bibitem{Zimdahl:2000zm}
W.~Zimdahl, D.~J.~Schwarz, A.~B.~Balakin and D.~Pavon,
``Cosmic anti-friction and accelerated expansion,''
Phys. Rev. D \textbf{64}, 063501 (2001)
doi:10.1103/PhysRevd.64.063501
[arXiv:astro-ph/0009353 [astro-ph]].

\bibitem{Kinnersley:1969zz}
W.~Kinnersley,
``Field of an Arbitrarily Accelerating Point Mass,''
Phys. Rev. \textbf{186}, 1335-1336 (1969)
doi:10.1103/PhysRev.186.1335

\bibitem{Kinnersley:1970zw}
W.~Kinnersley and M.~Walker,
``Uniformly accelerating charged mass in general relativity,''
Phys. Rev. D \textbf{2}, 1359-1370 (1970)
doi:10.1103/PhysRevD.2.1359

\bibitem{Bonnor:1994ir}
W.~M.~Bonnor,
``The Photon rocket,''
Class. Quant. Grav. \textbf{11}, 2007-2012 (1994)
doi:10.1088/0264-9381/11/8/008

\bibitem{Damour:1994dy}
T.~Damour,
``Photon rockets and gravitational radiation,''
Class. Quant. Grav. \textbf{12}, 725-738 (1995)
doi:10.1088/0264-9381/12/3/011
[arXiv:gr-qc/9412063 [gr-qc]].

\bibitem{Dain:1996eq}
S.~Dain, O.~M.~Moreschi and R.~J.~Gleiser,
``Photon rockets and the Robinson-Trautman geometries,''
Class. Quant. Grav. \textbf{13}, 1155-1160 (1996)
doi:10.1088/0264-9381/13/5/026
[arXiv:gr-qc/0203064 [gr-qc]].

\bibitem{Bonnor:1997jc}
W.~B.~Bonnor and M.~S.~Piper,
``The Gravitational wave rocket,''
Class. Quant. Grav. \textbf{14}, 2895-2904 (1997)
doi:10.1088/0264-9381/14/10/015
[arXiv:gr-qc/9702005 [gr-qc]].

\bibitem{Podolsky:2008wn}
J.~Podolsky,
``Photon rockets in (anti-)de Sitter universe,''
Phys. Rev. D \textbf{78}, 044029 (2008)
doi:10.1103/PhysRevD.78.044029
[arXiv:0806.2966 [gr-qc]].

\bibitem{Podolsky:2010ck}
J.~Podolsky,
``Photon rockets moving arbitrarily in any dimension,''
Int. J. Mod. Phys. D \textbf{20}, 335-360 (2011)
doi:10.1142/S0218271811018846
[arXiv:1006.1583 [gr-qc]].

\bibitem{Ge:2011ey}
H.~Ge, M.~Luo, Q.~Su, D.~Wang and X.~Zhang,
``Bondi-Sachs metrics and Photon Rockets,''
Gen. Rel. Grav. \textbf{43}, 2729-2742 (2011)
doi:10.1007/s10714-011-1197-3
[arXiv:1105.3258 [gr-qc]].

\bibitem{Hogan:2020duo}
P.~A.~Hogan and D.~Puetzfeld,
``Kerr analogue of Kinnersley\textquoteright{}s field of an arbitrarily 
accelerating point mass,''
Phys. Rev. D \textbf{102}, no.4, 044044 (2020)
doi:10.1103/PhysRevD.102.044044
[arXiv:2007.04685 [gr-qc]].

\bibitem{Forward84} R. L. Forward, ``Roundtrip interstellar travel using 
laser-pushed light-sails'', {\em J. Spacecraft Rockets} {\bf 21}, 187 
(1984).

\bibitem{Fuzfa:2019djg} A.~F\"uzfa, ``Interstellar travels aboard 
radiation-powered rockets,'' Phys. Rev. D \textbf{99}, no.10, 104081 
(2019) doi:10.1103/PhysRevD.99.104081 [arXiv:1902.03869 [gr-qc]].

\bibitem{Fuzfa:2020dgw}
A.~F\"uzfa, W.~Dhelonga-Biarufu and O.~Welcomme,
``Sailing Towards the Stars Close to the Speed of Light,''
Phys. Rev. Res. \textbf{2}, no.4, 043186 (2020)
doi:10.1103/PhysRevResearch.2.043186
[arXiv:2007.03530 [physics.pop-ph]].

\bibitem{Zeldovich:1970si} Y.~B.~Zeldovich, ``Particle production in 
cosmology,'' Pisma Zh. Eksp. Teor. Fiz. \textbf{12}, 443-447 (1970).

\bibitem{Hu} B. L. Hu, ``Vacuum viscosity description of quantum 
processes in the early universe'', Phys. Lett. {\bf 90A}, 375 (1982).

\bibitem{Schwarz:2001cf}
D.~J.~Schwarz, W.~Zimdahl, A.~B.~Balakin and D.~Pavon,
``Cosmic acceleration from effective forces?,''
doi:10.1007/10856495\_84
[arXiv:astro-ph/0110296 [astro-ph]].

\bibitem{Zimdahl:1996cg}
W.~Zimdahl, J.~Triginer and D.~Pavon,
``Collisional equilibrium, particle production and the inflationary 
universe,''
Phys. Rev. D \textbf{54}, 6101-6110 (1996)
doi:10.1103/PhysRevD.54.6101
[arXiv:gr-qc/9608038 [gr-qc]].

\bibitem{Zimdahl:1997qe}
W.~Zimdahl,
``Cosmological particle production and generalized thermodynamic 
equilibrium,''
Phys. Rev. D \textbf{57}, 2245-2254 (1998)
doi:10.1103/PhysRevD.57.2245
[arXiv:gr-qc/9711081 [gr-qc]].

\bibitem{ZimdahlBalakin98a} W. Zimdahl and A. B. Balakin, ``Kinetic theory 
for nongeodesic particle motion: self-interacting equilibrium states and 
effective viscous fluid pressures'', {\em Class. 
Quantum Grav.} {\bf 15}, 3259 (1998).

\bibitem{Zimdahl:1998zq}
W.~Zimdahl and A.~B.~Balakin,
``Inflation in a self-interacting gas universe,''
Phys. Rev. D \textbf{58}, 063503 (1998)
doi:10.1103/PhysRevD.58.063503
[arXiv:astro-ph/9809002 [astro-ph]].

\bibitem{Jordan38} P. Jordan, ``Zur empirischen kosmologie'', Naturwiss. 
\textbf{26}, 417 (1938).

\bibitem{Jordan:1959eg} P.~Jordan, ``The present state of Dirac's 
cosmological hypothesis,''
Z. Phys. \textbf{157}, 112-121 (1959) doi:10.1007/BF01375155

\bibitem{Brans:1961sx} C.~Brans and R.~H.~Dicke, ``Mach's principle and a 
relativistic theory of gravitation'', Phys. Rev. \textbf{124}, 925-935 
(1961) doi:10.1103/PhysRev.124.925.

\bibitem{Bergmann:1968ve} P.~G.~Bergmann, ``Comments on the scalar 
tensor theory'', Int. J. Theor. Phys. \textbf{1}, 25-36 (1968) 
doi:10.1007/BF00668828.

\bibitem{Nordtvedt:1968qs} K.~Nordtvedt, ``Equivalence Principle for 
Massive Bodies. 2. Theory'', Phys. Rev. \textbf{169}, 1017-1025 
(1968). doi:10.1103/PhysRev.169.1017.

\bibitem{Wagoner:1970vr} R.~V.~Wagoner, ``Scalar tensor theory and 
gravitational waves'', Phys. Rev. D \textbf{1}, 3209-3216 (1970) 
doi:10.1103/PhysRevD.1.3209.

\bibitem{Nordtvedt:1970uv} K.~Nordtvedt, Jr., ``PostNewtonian metric 
for a general class of scalar tensor gravitational theories and 
observational consequences'', Astrophys. J. \textbf{161}, 1059-1067 
(1970) doi:10.1086/150607.

\bibitem{Dicke:1961gz}
R.~H.~Dicke,
``Mach's principle and invariance under transformation of units,''
Phys. Rev. \textbf{125}, 2163-2167 (1962)
doi:10.1103/PhysRev.125.2163

\bibitem{Faraoni:2004pi}
V.~Faraoni, {\em Cosmology in Scalar-Tensor 
Gravity} (Kluwer Academic, Dordrecht, 2004), doi:10.1007/978-1-4020-1989-0

\bibitem{Faraoni:2023hwu}
V.~Faraoni and J.~Houle,
``More on the first-order thermodynamics of scalar-tensor and Horndeski 
gravity,''
Eur. Phys. J. C \textbf{83}, no.6, 521 (2023)
doi:10.1140/epjc/s10052-023-11712-7
[arXiv:2302.01442 [gr-qc]].

\bibitem{Taylor:1988nw}
T.~R.~Taylor and G.~Veneziano,
``Dilaton Couplings at Large Distances,''
Phys. Lett. B \textbf{213}, 450-454 (1988)
doi:10.1016/0370-2693(88)91290-7

\bibitem{Damour:1994zq}
T.~Damour and A.~M.~Polyakov,
``The String dilaton and a least coupling principle,''
Nucl. Phys. B \textbf{423}, 532-558 (1994)
doi:10.1016/0550-3213(94)90143-0
[arXiv:hep-th/9401069 [hep-th]].

\bibitem{Gasperini:1999ne}
M.~Gasperini,
``On the response of gravitational antennas to dilatonic waves,''
Phys. Lett. B \textbf{470}, 67-72 (1999)
doi:10.1016/S0370-2693(99)01309-X
[arXiv:gr-qc/9910019 [gr-qc]].
 
\bibitem{Faraoni:2018qdr} V.~Faraoni and J.~C\^ot\'e, ``Imperfect fluid 
description of modified gravities,'' Phys. Rev. D \textbf{98} no.~8, 
084019 (2018) doi:10.1103/PhysRevD.98.084019 [arXiv:1808.02427 [gr-qc]].

\bibitem{Faraoni:2021lfc} V.~Faraoni and A.~Giusti, ``Thermodynamics of 
scalar-tensor gravity,'' Phys. Rev. D \textbf{103}, no.12, L121501 (2021) 
doi:10.1103/PhysRevD.103.L121501 [arXiv:2103.05389 [gr-qc]].

\bibitem{Faraoni:2021jri} V.~Faraoni, A.~Giusti and A.~Mentrelli, ``New 
approach to the thermodynamics of scalar-tensor gravity,'' Phys. Rev. D 
\textbf{104}, no.12, 124031 (2021) doi:10.1103/PhysRevD.104.124031 
[arXiv:2110.02368 [gr-qc]].

\bibitem{Giusti:2022tgq}A.~Giusti, S.~Giardino and 
V.~Faraoni,``Past-directed scalar field gradients and scalar-tensor 
thermodynamics,''Gen. Rel. Grav. \textbf{55}, no.3, 47 
(2023)doi:10.1007/s10714-023-03095-7[arXiv:2210.15348 [gr-qc]].

\bibitem{Quiros:2019gai} U.~Nucamendi, R.~De Arcia, T.~Gonzalez, 
F.~A.~Horta-Rangel and I.~Quiros, ``Equivalence between Horndeski and 
beyond Horndeski theories and imperfect fluids,'' Phys. Rev. D 
\textbf{102} (2020) no.8, 084054, doi:10.1103/PhysRevD.102.084054 
[arXiv:1910.13026 [gr-qc]].

\bibitem{Giusti:2021sku}A.~Giusti, S.~Zentarra, L.~Heisenberg and 
V.~Faraoni,``First-order thermodynamics of Horndeski gravity,''Phys. Rev. 
D \textbf{105}, no.12, 124011 
(2022)doi:10.1103/PhysRevD.105.124011[arXiv:2108.10706 [gr-qc]].

\bibitem{Eckart40} C.~Eckart, ``The thermodynamics of irreversible 
processes.~3. Relativistic theory of the simple fluid,'' Phys. Rev. 
\textbf{58} (1940), 919-924 doi:10.1103/PhysRev.58.919

\bibitem{Miranda:2022wkz}M.~Miranda, D.~Vernieri, S.~Capozziello and 
V.~Faraoni,``Fluid nature constrains Horndeski gravity,''Gen. Rel. Grav. 
\textbf{55}, no.7, 84 
(2023)doi:10.1007/s10714-023-03128-1[arXiv:2209.02727 [gr-qc]].

\bibitem{Giardino:2022sdv}S.~Giardino, V.~Faraoni and 
A.~Giusti,``First-order thermodynamics of scalar-tensor cosmology,''JCAP 
\textbf{04}, no.04, 053 
(2022)doi:10.1088/1475-7516/2022/04/053[arXiv:2202.07393 [gr-qc]].

\bibitem{Faraoni:2022gry}V.~Faraoni, S.~Giardino, A.~Giusti and 
R.~Vanderwee,``Scalar field as a perfect fluid: thermodynamics of 
minimally coupled scalars and Einstein frame scalar-tensor gravity,''Eur. 
Phys. J. C \textbf{83}, no.1, 24 
(2023)doi:10.1140/epjc/s10052-023-11186-7[arXiv:2208.04051 [gr-qc]].

\bibitem{Miranda:2022uyk}M.~Miranda, P.~A.~Graham and 
V.~Faraoni,``Effective fluid mixture of tensor-multi-scalar gravity,''Eur. 
Phys. J. Plus \textbf{138}, no.5, 387 
(2023)doi:10.1140/epjp/s13360-023-03984-5[arXiv:2211.03958 [gr-qc]].

\bibitem{Faraoni:2022doe}V.~Faraoni, A.~Giusti, S.~Jose and 
S.~Giardino,``Peculiar thermal states in the first-order thermodynamics of 
gravity,''Phys. Rev. D \textbf{106}, no.2, 024049 
(2022)doi:10.1103/PhysRevD.106.024049[arXiv:2206.02046 [gr-qc]].

\bibitem{Faraoni:2022jyd}V.~Faraoni and T.~B.~Fran\c{c}onnet,``Stealth 
metastable state of scalar-tensor thermodynamics,''Phys. Rev. D 
\textbf{105}, no.10, 104006 
(2022)doi:10.1103/PhysRevD.105.104006[arXiv:2203.14934 [gr-qc]].

\bibitem{Giardino:2023qlu}S.~Giardino, A.~Giusti and V.~Faraoni,``Thermal 
stability of stealth and de Sitter spacetimes in scalar-tensor 
gravity,''Eur. Phys. J. C \textbf{83}, no.7, 621 
(2023)doi:10.1140/epjc/s10052-023-11697-3[arXiv:2302.08550 [gr-qc]].

\bibitem{Faraoni:2022fxo}V.~Faraoni, P.~A.~Graham and 
A.~Leblanc,``Critical solutions of nonminimally coupled scalar field 
theory and first-order thermodynamics of gravity,''Phys. Rev. D 
\textbf{106}, no.8, 084008 
(2022)doi:10.1103/PhysRevD.106.084008[arXiv:2207.03841 [gr-qc]].

\bibitem{Giardino:2023sw} S.~Giardino and A.~Giusti, ``First-order 
thermodynamics of scalar-tensor gravity,'' Ricerche di Matematica (2023) 
doi:10.1007/s11587-023-00801-0 [arXiv:2306.01580 [gr-qc]].

\bibitem{Miranda:2024dhw}M.~Miranda, S.~Giardino, A.~Giusti and 
L.~Heisenberg,``First-order thermodynamics of Horndeski 
cosmology,''[arXiv:2401.10351 [gr-qc]].

\bibitem{Giardino:2023ygc}
S.~Giardino and A.~Giusti,
``First-order thermodynamics of scalar-tensor gravity,''
doi:10.1007/s11587-023-00801-0
[arXiv:2306.01580 [gr-qc]].

\bibitem{Houle:2024sxs}
J.~Houle and V.~Faraoni,
``New phenomenology in the first-order thermodynamics of scalar-tensor 
gravity for Bianchi universes,''
Phys. Rev. D \textbf{110}, no.2, 024067 (2024)
doi:10.1103/PhysRevD.110.024067
[arXiv:2404.19470 [gr-qc]].

\bibitem{Karolinski:2024nwp}
N.~Karolinski and V.~Faraoni,
``Einstein gravity as the thermal equilibrium state of a nonminimally 
coupled scalar field geometry,''
Phys. Rev. D \textbf{109}, no.8, 084042 (2024)
doi:10.1103/PhysRevD.109.084042
[arXiv:2401.04877 [gr-qc]].

\bibitem{Giardino:2024pla}
S.~Giardino,
``First-order thermodynamics of modified gravity,''
doi:10.11588/heidok.00035192


\end{thebibliography}


\end{document}